\documentclass[12pt]{article}
\usepackage{amsfonts,amsmath,amssymb,amscd,mathrsfs,comment,enumitem,graphicx, color}

\usepackage{cite}

\DeclareMathAlphabet{\mathpzc}{OT1}{pzc}{m}{it}

\title{The Bohmian solution to the problem of time}

\author{Ward Struyve\footnote{Instituut voor Theoretische Fysica, KU Leuven, Belgium}$^{*}$\footnote{Centrum voor Logica en Filosofie van de Wetenschappen, KU Leuven, Belgium}  }

\date{}

\def\de{\delta}

\def\ka{\kappa}

\def\pa{\partial}

\def\ka{\kappa}
\def\ii{\textrm i}
\def\ee{\textrm e}

\newcommand{\be}{\begin{equation}}
\newcommand{\en}{\end{equation}}
\newcommand{\bi}{\begin{itemize}}
\newcommand{\ei}{\end{itemize}}

\begin{document}
\maketitle

\begin{abstract}
  \noindent

In canonical quantum gravity the wave function of the universe is static, leading to the so-called problem of time. We summarize here how Bohmian mechanics solves this problem. 

\end{abstract}

\noindent
{\em This paper is dedicated to Detlef D\"urr with whom I had the pleasure to discuss quantum gravity and the problem of time on many occasions. For Detlef conceptual clarity was always quintessential to get a firm grip on this problem.} \\

\section{Introduction}

Canonical quantum gravity is an approach to quantum gravity which is obtained by applying the usual quantization techniques to the classical theory of gravity, general relativity \cite{kiefer07}. These techniques have been successfully applied before to classical Yang-Mills theories to construct quantum theories describing the fundamental interactions other than gravity, leading to the Standard Model of elementary particle physics, and the hope is that they lead to similar success in the case of gravity. As usual, the quantization proceeds by bringing general relativity into Hamiltonian form and by replacing classical variables by operators acting on wave functions. This results in a set of wave equations for the wave function, one of which is known as the Wheeler-DeWitt equation. There is a host of problems with these wave equations, of both technical and conceptual nature. On the technical level, the main problem is that the wave equations are merely defined formally. Since the theory is non-renormalizable (unlike Yang-Mills theories), this problem cannot be bypassed by the usual perturbation methods. On the conceptual level, there is the measurement problem, which carries over from ordinary quantum theory, and which is especially severe in this context since the aim is to apply the theory to the whole universe, so that there are no outside observers performing measurements. Another immediate problem is the problem of time. The wave equations entail a static wave function, which hence seems unable to describe our changing universe. Various ideas have been proposed on how to overcome this problem \cite{kuchar92,isham92,kiefer07,anderson17}, without any consensus about the right approach. A recurring issue with many of these proposals is the lack of ontological clarity. What is the ontology in quantum gravity? What is the role of the wave function? Is it part of the ontology? Or merely a statistical representation of some sort? Does space-time itself exists? Or is it emergent?  What are the objects in space-time?{\footnote{The wave function itself is not an object in space-time, but in some Hilbert space.}} A proper resolution of this problem requires clarity about these questions.

In Bohmian mechanics the ontology is clear. Non-relativistic Bohmian mechanics is about point-particles in Galilean space-time moving with a velocity that depends on the wave function which satisfies the usual Schr\"odinger equation \cite{bohm93,holland93b,duerr09}. In Bohmian quantum gravity \cite{goldstein04,pinto-neto05a,pinto-neto19}, space-time itself is dynamical, like in general relativity. As in the classical Hamiltonian formulation, it can be written as a geometrodynamics, i.e., as a dynamics of a spatial 3-metric, whose time evolution determines a space-time metric. This time evolution, together with that of the matter degrees of freedom (particles or fields), is determined by the wave function which satisfies the wave equations of canonical quantum gravity. Even though the wave function is static, the spatial 3-metric and the matter degrees of freedom generically change over time. As such, the problem of time is immediately solved. 
 
Before turning to Bohmian quantum gravity, it is instructive to consider the quantization of the non-relativistic particle. By putting the dynamics in parameterized form, it shares many features with the gravitational case, see e.g.\ \cite{kuchar93,ruffini95,rovelli14}. We will give some details of the classical Hamiltonian formulation and how quantization leads to a stationary wave function. This also allows us to illustrate that classically there is no problem of time, contrary to what is sometimes claimed, see e.g.\ \cite{earman02a} and rebuttals \cite{maudlin02b,healey02}.

\section{Non-relativistic particle}\label{particle}
Consider a single particle in Galilean space-time, whose possible trajectories ${\bf x}(t)$ satisfy the Newtonian equation 
\be
m \ddot {\bf x}(t) = - {\boldsymbol \nabla} V({\bf x}(t)),
\label{1}
\en 
with $V({\bf x})$ a potential. The dynamics can be expressed in other ways, for example in the form of an action principle, with action $S = \int dt L$ and Lagrangian
\be
L = \frac{m}{2} \dot {\bf x}^2 - V({\bf x}).
\en
Extrema of the action satisfy \eqref{1}. From the Lagrangian, also the Hamilton formulation can be obtained. With momentum ${\bf p} = \pa L/ \pa \dot {\bf x} = m\dot {\bf x}$, the Hamiltonian is 
\be
H = \frac{1}{2m}{\bf p}^2 + V
\label{5}
\en
and Hamilton's equations read
\be
\dot {\bf x} = \frac{{\bf p}}{m} , \qquad \dot {\bf p} = - {\boldsymbol \nabla}  V.
\en
Of course, by casting the dynamics this way, the ontology has not changed. The theory is still about a particle moving in space-time. It is not about, say, a point moving in phase space. 

Another way to formulate the dynamics is by considering a different parameterization of the trajectory. Rather than parameterizing it by the time $t$, a parameter $s$ can be introduced so that the trajectories are represented by $(t(s),{\bf x}(s))$. Denoting the derivatives with respect to $s$ by primes, the dynamics reads
\be 
m \frac{1}{t'(s)} \left( \frac{1}{t'(s)} {\bf x}'(s) \right)' = - {\boldsymbol \nabla}  V({\bf x}(s)).
\label{para}
\en
This dynamics has a reparameterization symmetry. Namely, for any solution $(t(s),{\bf x}(s))$, also $(t({\tilde s}(s)),{\bf x}({\tilde s}(s)))$ will be a solution, for any different parameterization ${\tilde s}(s)$. But any two such solutions represent the same physical situation. Namely, they represent the same curve in space-time. The parameterization is just a part of the mathematical representation, without any physical significance. It is an instance of a gauge symmetry. 

This symmetry is connected with indeterminism. Specifying ${\bf x}(t)$ and $\dot {\bf x}(t)$ at a particular time determines a unique solution to \eqref{1}. On the other hand, specifying $(t(s),{\bf x}(s))$ at a certain value for $s$ does not yield a unique solution to \eqref{para}, which is an immediate consequence of the reparameterization symmetry. For a Lagrangian theory, indeterminism is often seen as the defining characteristic of gauge \cite{sundermeyer82,henneaux92}. 

Also the parameterized dynamics can be formulated in Hamiltonian form. Starting from the Lagrangian 
\be
{\bar L} = t'\left( \frac{m}{2} \frac{{\bf x}'^2}{t'^2} - V({\bf x}) \right),
\en
the conjugate momenta are 
\be
{\bf p} = \frac{\pa {\bar L}}{\pa {\bf x}' } =  m \frac{{\bf x}'}{t'} , \qquad p_t = \frac{\pa {\bar L}}{\pa t' } = - \frac{m}{2} \frac{{\bf x}'^2}{t'^2} - V({\bf x}) .
\label{20}
\en
These momenta imply the constraint
\be
C = p_t + \frac{1}{2m}{\bf p}^2 + V =  0,
\en
which implies that the relations \eqref{20} cannot be inverted to yield the velocities in terms of the momenta. This means we have to resort to the formalism of constrained Hamiltonian dynamics \cite{sundermeyer82,henneaux92}. The canonical Hamiltonian vanishes,
\be
{\bar H} = p_t t' + {\bf p}{\bf x}' - {\bar L} = 0, 
\en
but Hamilton's equations are derived from the total Hamiltonian
\be
H_T = {\bar H} + N C =  N C, 
\label{23}
\en
where $N$ is an arbitrary (non-zero) function of the phase space variables. Together with the constraint $C=0$, Hamilton's equation are 
\be
t' = N, \qquad p_t' = 0, \qquad  {\bf x}' = N \frac{\bf p}{m} ,\qquad {\bf p}' =  - N {\boldsymbol \nabla}  V.
\en
Because of the arbitrariness of $N$, the parameterization invariance has become explicit. Again, the dynamics was recast in a different way, without affecting the possible trajectories in space-time.

Another formulation of interest is the one in terms of the reduced phase-space  \cite{sundermeyer82,henneaux92}. The reduced phase-space is obtained by eliminating the gauge variables. The gauge variables are identified as those variables that evolve completely freely. They are contrasted with what are sometimes called the {\em observables} \cite{henneaux92} or {\em true degrees of freedom} \cite{sundermeyer82}, which evolve deterministically. In the present case, this means that the latter have zero Poisson brackets with the constraint and hence also that these variables must be static. Does this mean there is no change, as some have argued? This will depend on the ontological significance one attaches to those variables. Let us first make this formulation concrete. In general, the reduced phase space is hard to find, but in the case of $V=0$, it is obtained by performing the following canonical transformation
\be
{\tilde {\bf x}} = {\bf x} -\frac{1}{m} {\bf p}t , \qquad {\tilde t} = t , \qquad {\tilde {\bf p}} = {\bf p},  \qquad {\tilde p}_{\tilde t} =  p_t + \frac{1}{2m}{\bf p}^2.
\en
Along a trajectory $(t(s),{\bf x}(s))$, ${\tilde {\bf x}}$ and ${\tilde {\bf p}}$ correspond to the position and momentum at $t=0$. In terms of the new variables, the constraint reads ${\tilde p}_{\tilde t} = 0$ and the Hamiltonian is
\be
{\tilde H} = N {\tilde p}_{\tilde t}.
\en
The equations of motion are
\be
 {\tilde t}' = N , \qquad {\tilde p}_{\tilde t} = 0, \qquad {\tilde {\bf x}}' = 0 , \qquad {\tilde {\bf p}}' = 0. 
\label{25}
\en
So the dynamics of ${\tilde {\bf x}}$ and ${\tilde {\bf p}}$ decouples from that of ${\tilde t}$ and ${\tilde p}_{\tilde t}$. Moreover, because of the arbitrariness of $N$, the dynamics of ${\tilde t}$ is completely free and is identified as the gauge degree of freedom. By dropping the variables ${\tilde t}$ and ${\tilde p}_{\tilde t}$ from the description, the reduced phase space is obtained, which is parameterized by ${\tilde {\bf x}}$ and ${\tilde {\bf p}}$. These variables ${\tilde {\bf x}}$ and ${\tilde {\bf p}}$ are just static. (After all, they correspond to the initial state.) What does all this buy us? The reduced phase space formulation is another way of formulating the dynamics of the Newtonian particle, where the unphysical parameterization has indeed been eliminated. But, as with the other formulations, the ontology has not changed by doing so. What is physically real is the particle whose position is changing over time according to ${\bf x}(t) =  {\tilde {\bf p}}t/m + {\tilde {\bf x}}$. The path can indeed be fully specified by the initial state $({\tilde {\bf x}},{\tilde {\bf p}})$. But that is not to say that the ontology should comprise just that initial state. One could entertain the latter possibility, but then indeed it becomes problematic to account for a changing world.{\footnote{An ontology based on the reduced phase space is not always problematic. In other cases, it makes sense to consider such an ontology. For example, in the case of the free electromagnetic field, the reduced phase space can be parameterized by the transverse part of the field potential, together with its conjugate momentum, and one could entertain an ontology based on this field.}}

To finish this section, a word about clock variables. Consider $n$ non-relativistic particles. Then a collection  of particles may serve as a clock whenever its configuration is non-stationary. As a concrete example, suppose there is a particle for which one of the coordinates $z(t)$ is monotonically increasing with time.{\footnote{Rather than considering the positions as functions of time $t$, we could equally well have done this discussion in terms of an arbitrary parameterization $s$. That there is an external time plays no role here, as long as there is change with respect to $s$.}} The motion of the other particles might then be expressed with respect to the clock variable $z$, by inverting the relation, i.e., $t(z)$, and substituting that in the positions of the other particles: $ {\bf x}_1(t(z)),\dots,{\bf x}_n(t(z))$. Other variables could act as clock variables, but it is always the dynamics of the variables that determines whether they can serve as clock variables or not, and how one can switch between clock variables.

\section{Non-relativistic Bohmian mechanics}
Canonical quantization is a recipe for getting a quantum theory from a classical theory, starting from the Hamiltonian formulation. In the previous section, we have seen three different Hamiltonian formulations. Let us now apply the quantization rules. Starting from the Hamiltonian \eqref{5} this results in the familiar Schr\"odinger equation{\footnote{Throughout we assume units such that $\hbar = c =1$.}} 
\be
\ii \pa_t \psi({\bf x},t) = -\frac{1}{2m} \nabla^2 \psi ({\bf x},t)+ V \psi({\bf x},t)
\label{60}
\en
for $\psi({\bf x},t)$. Starting from the Hamiltonian \eqref{23} yields
\be
\ii \pa_s \psi ({\bf x},t,s)=0 , \qquad - \ii \pa_t \psi ({\bf x},t,s)- \frac{1}{2m} \nabla^2 \psi ({\bf x},t,s)+ V \psi({\bf x},t,s) =0,
\label{61}
\en
for the wave function $\psi({\bf x},t,s)$. (The second equation arises here as the operator equivalent of the constraint $C=0$.) Here we see the appearance of a static wave function, i.e., static with respect to the parameterization $s$. As a wave equation, this is just the Schr\"odinger equation again.{\footnote{To show the equivalence as quantum theories, more work is needed, by also considering the associated Hilbert spaces.}} Starting from the reduced phase space, we get
\be
\ii \pa_s \psi({\tilde {\bf x}}) =0 ,
\label{62}
\en
for $\psi({\tilde {\bf x}})$. This wave equation is also equivalent with \eqref{60}, provided $\psi({\tilde {\bf x}})$ is taken as the initial wave function $\psi({\bf x},0)$ \cite{ruffini95}.

In the Bohmian theory, in addition to the wave function which satisfies the Schr\"odinger equation \eqref{60}, there is an actual point-particle with position ${\bf x}$ whose velocity is given by 
\be
\dot {\bf x} = {\bf v}^\psi({\bf x},t),
\label{70}
\en
where
\be
{\bf v}^\psi({\bf x},t) = \frac{1}{m}{\textrm{Im}} \frac{{\boldsymbol \nabla} \psi({\bf x},t)}{\psi({\bf x},t)}.
\en
The velocity field is of the form ${\bf v}^\psi = {\bf j}^\psi/ |\psi|^2$, where ${\bf j}^\psi$ is the usual current, satisfying the continuity equation
\be
\pa_t |\psi|^2 + {\boldsymbol \nabla} \cdot {\bf j}^\psi = 0.
\en
This form of the velocity field can be used to formulate a Bohmian dynamics in other contexts \cite{struyve09a}. For example, in the formulation \eqref{61}, we can consider trajectories in parameterized form $x(s) = (t(s),{\bf x}(s))$, with a velocity determined by the current $j^\psi=(|\psi|^2,{\bf j}^\psi)$. Since the parameterization of the curve is arbitrary, the dynamics can be written as
\be
x' \sim j^\psi,
\en
which geometrically expresses that the tangent vector to the curve in space-time is always proportional to $j^\psi$. We can explicitly introduce an arbitrary (non-vanishing) factor $N(x,s)$ and write the dynamics as
\be
x' = \frac{N}{|\psi|^2} j^\psi
\en
or
\be
t' = N, \qquad  {\bf x}' = N {\bf v}^\psi({\bf x},t).
\en
It is clear that this is the dynamics \eqref{70} written in parameterized form. In addition, even though the wave function is static with respect to $s$, the actual configuration $x$ generically is not. (The fact that $\psi$ still depends on $t$ does not matter, see further below.) Applying the same recipe to the wave equation \label{62} arising from reduced phase space quantization, suggest a Bohmian dynamics ${\tilde {\bf x}}' = 0$. This would only make sense if ${\tilde {\bf x}}$ is taken as the initial position with trajectories again given by \eqref{70}.

Before turning to gravity, we want to further illustrate some aspects of the non-relativistic Bohmian dynamics that will be relevant in dealing with the problem of time in the context of quantum gravity. 

Consider an $n$-particle system, with particle positions ${\bf x}_k$, $k=1,\dots,n$, satisfying the guidance equations
\be
\dot {\bf x}_k =  {\bf v}_k^\psi = \frac{1}{m_k}{\textrm{Im}} \frac{{\boldsymbol \nabla}_k \psi}{\psi},
\en
with $\psi({\bf x}_1,\dots, {\bf x}_n,t)$ now satisfying the many particle Schr\"odinger equation
\be
\ii \pa_t \psi = -\frac{1}{2m_k} \sum_k \nabla^2_k \psi + V \psi.
\label{60.2}
\en
The first observation is that for a stationary state $\psi=\ee^{-\ii Et}\phi({\bf x}_1,\dots, {\bf x}_n)$, the time dependence of the wave function is trival. For such a wave function, the velocity field ${\bf v}_k^\psi$ is time-independent, but still the trajectories could be highly non-trivial (see e.g.\ the simulations in \cite{struyve16}). 

Second, even if the wave function of the total system is stationary, the wave function of a subsystem will typically be non-stationary \cite{duerr92a,goldstein04,duerr09}. To see this, we first need to define the subsystem wave function. Consider the wave function $\psi({\bf x}_1,\dots,{\bf x}_n,{\bf y}_1,\dots,{\bf y}_m,t)$, where the $x$-coordinates describe the subsystem and the $y$-coordinates its environment. The wave function of the subsystem --- called the {\em conditional wave function} --- can then be defined as 
\be
\psi_s ({\bf x}_1,\dots,{\bf x}_n,t) = \psi({\bf x}_1,\dots,{\bf x}_n,{\bf y}_1(t),\dots,{\bf y}_m(t),t),
\en
which is the total wave function evaluated for the actual particle positions ${\bf y}_1(t),\dots,{\bf y}_m(t)$. One of the reasons this definition is natural is that the velocities of the particles of the subsystem can be written either in terms of $\psi$ or $\psi_s$, i.e.,
\be
\dot {\bf x}_k =   \frac{1}{m_k}{\textrm{Im}} \frac{{\boldsymbol \nabla}_k \psi}{\psi} =   \frac{1}{m_k}{\textrm{Im}} \frac{{\boldsymbol \nabla}_k \psi_s}{\psi_s}.
\en
The conditional wave function $\psi_s$ will generically evolve in time and under certain conditions it will satisfy an effective Schr\"odinger equation.

\section{Quantum gravity}
General relativity describes how space-time, described by a manifold ${\mathcal M}$ and metric field $g_{\mu \nu}(x)$, interacts with matter. The equations of motion are given by the Einstein field equations, together with equations for the matter (say particles or a field). The theory exhibits a gauge invariance, namely the invariance under coordinate transformations, i.e., invariance under space-time diffeomorphisms. Two metrics connected by a diffeomorphism are considered physically equivalent and they are said to determine the same 4-geometry. 

When a space-time is globally hyperbolic, it admits a foliation of ${\mathcal M}$ in terms of spatial hypersurfaces, because ${\mathcal M}$ then topologically equals ${\mathbb R} \times \Sigma$, with $\Sigma$ a 3-surface. This allows a splitting of space-time into space and time, which is required for the Hamiltonian formulation.{\footnote{Just like in special relativity, this splitting is not unique.}} Coordinates $x^\mu=(t,{\bf x})$ can be chosen such that the coordinate $t$ labels time and ${\bf x}$ are coordinates on $\Sigma$. In terms of these coordinates, the space-time metric can be written as
\be
g_{\mu \nu}=
\begin{pmatrix}
	N^2 - N_i N^i & -N_i \\
	-N_i & - h_{ij}
\end{pmatrix} ,
\label{g1}
\en
where $N > 0$ is the lapse function, $N_i = h_{ij}N^j$ are the shift functions, and $h_{ij}$ is the induced Riemannian metric on the leaves of the foliation. In the Hamiltonian formulation, only the evolution of the 3-metric $h_{ij}$ is non-trivial; the evolution of the lapse and shift functions is completely undetermined. This dynamics is referred to as {\em geometrodynamics}.

Canonical quantization of the theory leads to the following wave equation for the wave functional $\Psi(h_{ij},\phi)$, where $h_{ij}$ is again the 3-metric and $\phi$ is a scalar field. (Rather than having a scalar field describing the matter, there are of course other possibilities.) The wave functional satisfies the functional Schr\"odinger equation{\footnote{We used the same operator ordering as in \cite{duerr20b}.}}
\be
\ii \pa_t \Psi = \int d^3 x \left(N {\widehat {\mathcal H}} + N^i {\widehat {\mathcal H}}_i  \right) \Psi,
\label{cqg10}
\en
where
\be
{\widehat {\mathcal H}}   =  - \ka {\sqrt h}\frac{\delta}{\delta h_{ij}} \left( \frac{1}{\sqrt h} G_{ijkl}\frac{\delta}{\delta h_{kl}}\right) - \frac{1}{2}\frac{1}{\sqrt h}\frac{\delta^2}{\delta \phi^2} + {\mathcal V}(h,\phi), 
\label{q3}
\en
\be
{\widehat {\mathcal H}}_i  =  -2 h_{ik}D_j\frac{\delta }{\delta h_{jk}} + \frac{1}{2} \left( \pa_i \phi \frac{\de }{\de \phi} + \frac{\de }{\de \phi} \pa_i \phi\right),
\label{q3.1}
\en
with $\kappa = 16\pi G$, $G$ the gravitational constant, $G_{ijkl}$ the DeWitt metric (which depends on the 3-metric), $h$ the determinant of $h_{ij}$,  and ${\mathcal V}$ is a potential. In addition, the wave functional satisfies the constraints:
\be
{\widehat {\mathcal H}}({\bf x}) \Psi = 0 , \qquad  {\widehat {\mathcal H}}_i ({\bf x})\Psi = 0, \quad i=1,2,3 ,
\label{cqg12}
\en
the first of which is the Wheeler-DeWitt equation. These constraints immediately imply that 
\be
\pa_t \Psi = 0 ,
\en
so that $\Psi$ must be time-independent. This is the source of the problem of time.

In Bohmian quantum gravity, there is an actual 3-metric $h_{ij}$ and a field $\phi$, evolving according to
\be
{\dot{h}}_{ij} = 2\ka NG_{ijkl}\frac{\delta S}{\delta h_{kl}} + D _i N_j + D _j N_i,
\label{bcqg1}
\en
\be
\dot \phi =   \frac{N}{\sqrt h} \frac{\delta S}{\delta \phi} + N^i \pa_i \phi,
\label{dotphi}
\en
with $\Psi=|\Psi|\ee^{\ii S}$ and $D_i$ the 3-dimensional covariant derivative. Given a lapse and a shift function, a solution for the 3-metric defines a 4-metric using \eqref{g1}. There is an important difference with classical geometrodynamics, however. Whereas classically, any choice of lapse or shift defines the same 4-geometry, this is no longer the case for Bohmian geometrodynamics: different choices for the lapse function will lead to different 4-geometries (unless the difference is just a $t$-dependent factor). Different shift functions do not affect the 4-geometry. This implies that the theory is not invariant under general diffeomorphisms, but only under spatial diffeomorphisms. This is akin to the situation in special relativity, where the simplest Bohmian formulations employ a preferred reference frame (or more generally a preferred space-time foliation), which breaks the Lorentz symmetry (but not on the observational level). The reason is that the non-locality which is inherent to quantum theory (due to Bell's theorem) is hard to marry with Lorentz invariance, or with diffeomorphism invariance in the case of Bohmian quantum gravity. For some suggestions of how to formulate Lorentz invariant Bohmian theories, see \cite{duerr14a}.

There is no problem of time in Bohmian quantum gravity \cite{vink92,goldstein04,pinto-neto19}. While the wave function is static, the 3-metric and the scalar field generically evolve in time. (For real wave functions, i.e.\ $S=0$, such as the Hartle-Hawking wave function, there is no motion, and hence from the Bohmian point of view this wave function is inadequate to describe our universe \cite{squires92b,shtanov96}.) Unlike the wave function of the universe, the wave function of a subsystem is generically time dependent and in certain cases it satisfies a Schr\"odinger equation. For example, in the study of cosmological perturbations (see \cite{pinto-neto19} and references therein), relevant in for example cosmological inflation theory, one can consider a decomposition of the metric and the scalar field in terms of a homogeneous and isotropic background component and fluctuations: $h = h_0 + h_1$, $\phi=\phi_0 + \phi_1$. If the wave function is approximately of the form $\Psi(h,\phi) = \Psi_0(h_0,\phi_0) \Phi(h_0,\phi_0,h_1,\phi_1)$, where $\Psi_0$ yields the dominant contribution to the velocity field for $h_0$ and $\phi_0$, the conditional wave function for the perturbations $h_1$ and $\phi_1$  might satisfy (given again some approximations) some effective Schr\"odinger equation with a Hamiltonian depending on the actual Bohmian configuration $(h_0(t),\phi_0(t))$ (which is usually taken to be classical).

There is an approach to the problem of time that is very close to the Bohmian one \cite{vink92,callender96,goldstein04}. In that approach, there is no time at the fundamental level, but it is said to emerge in a semi-classical approximation (see \cite{kiefer07} for details and the history of this approach, see also \cite{kuchar92,chua21} for critical assessments). The starting point is to consider an approximation
\be
\Psi(h_{ij},\phi) \approx \Psi_0(h_{ij})\Phi(h_{ij},\phi),
\en
where $\Psi_0$ is approximately a ``classical state'', i.e., it can be approximated by the dominating contribution in its WKB expansion. Next, a classical trajectory $h_{ij}(t)$ is considered, determined by \eqref {bcqg1}, where the phase is that of $\Psi_0$. (That the trajectory is indeed classical stems from the assumptions about $\Psi_0$.) Then the wave function 
\be
\psi(\phi,t) = \Phi(h_{ij}(t),\phi)
\label{con}
\en
is defined and shown to approximately satisfy the time-dependent Schr\"odinger equation
\be
\ii \pa_t \psi = \int d^3 x \left(N {\widehat {\mathcal H}}_M + N^i {\widehat {\mathcal H}}_{Mi}  \right) \psi,
\label{con2}
\en
where ${\widehat {\mathcal H}}_M$ and ${\widehat {\mathcal H}}_{Mi}$ are the matter part of respectively \eqref{q3} and \eqref{q3.1}, evaluated for $h_{ij}(t)$. This is the Schr\"odinger equation for a matter field in the external classical metric $h_{ij}(t)$. Granting the approximations, the introduction of the classical trajectory and of the wave function $\psi$ is rather ad hoc and has no ontological basis. However, what is outlined here can be perfectly justified from the Bohmian point of view. The wave function $\psi$ in \eqref{con} then basically amounts to the conditional wave function and in the case the Bohmian field $h_{ij}$ evolves approximately classical, with a velocity field approximately determined solely by $\Psi_0$, the conditional wave function for the matter field will approximately satisfy \eqref{con2}. Unlike the postulation of \eqref{con} above, there is nothing ad hoc about the conditional wave function in Bohmian mechanics. It is the wave function that can be used to write the velocity field of the scalar field. So given that the approximations are justified (and the Bohmian theory makes precise the conditions under which they are), the wave function $\psi$ can be used to write the velocity and hence the dynamics of the actual field $\phi(t)$. 

Actually, as was noted before by Padmanabhan \cite{padmanabhan90} and Greensite \cite{greensite90,greensite91}, the assumption of classicality of $\Psi_0$ played no crucial role in the above analysis. That is, the analysis can be generalized to other wave functions $\Psi$, by considering a trajectory $(h_{ij}(t),\phi(t))$ satisfying \eqref{bcqg1} and \eqref{dotphi}, instead of a classical trajectory, and by replacing \eqref{con} by the wave function $\Psi$ evaluated for one of the degrees of freedom of the metric $h_{ij}(t)$. Clearly, this amounts to just assuming the Bohmian dynamics and employing the conditional wave function to get a time-dependent wave equation.{\footnote{The papers of Padmabhan and Greensite actually predate the development of Bohmian quantum gravity, which was initiated shortly afterwards by Vink \cite{vink92}. Vink also emphasized that Bohmian quantum gravity yields a generalization of the semi-classical approach to time.}} 
In any case, this generalization allows for a much larger class of approximations than merely the one obtained by assuming classicality of $\Psi_0$ (see sections 7 and 9 of \cite{pinto-neto19} and references therein).

There are also approaches to the problem of time that proceed by just postulating clock variables, see e.g.\ \cite{page83}. However, by merely postulating them rather than deriving them from the ontology, one tends to commit redundancy and risk inconsistency \cite{bell82}. At the end of section \ref{particle}, it was discussed how clock variables can be defined in the context of classical mechanics. In Bohmian mechanics they are defined exactly the same way and as such this could provide an underpinning of these clock-based approaches to the problem of time.

We have only discussed canonical quantum gravity. Loop quantum gravity \cite{thiemann07,rovelli14}, which is also obtained by quantizing general relativity, but based on a different representation than canonical quantum gravity, also suffers the problem of time. Here too, a Bohmian version could solve the problem. See \cite{struyve17b} for the treatment of mini-superspace models (which are simplified models of quantum gravity, potentially applicable in cosmology).

\section{Acknowledgments}
It is a pleasure to thank Detlef D\"urr, Sheldon Goldstein, Stefan Teufel, Roderich Tumulka, Hans Westman and Nino Zangh\`i for discussions on this topic over the years. This work was presented at the OLOFOS Seminar in Leuven. I thank the audience and especially the respondent Alexandre Guay for the discussions. This work is supported by the Research Foundation Flanders (Fonds Wetenschappelijk Onderzoek, FWO), Grant No. G0C3322N.


\begin{thebibliography}{10}

\bibitem{kiefer07}
{C.\ Kiefer, {\em Quantum Gravity}, second edition, Oxford University Press,
  Oxford (2007).}

\bibitem{kuchar92}
{K.V.\ Kucha{\u{r}}, ``Time and interpretations of quantum gravity'', in {\em
  Proceedings of the 4th Canadian Conference on General Relativity and
  Relativistic Astrophysics}, eds.\ G.\ Kunstatter, D.\ Vincent and J.\
  Williams, World Scientific, Singapore (1992), reprinted in {\em Int.\ J.\
  Mod.\ Phys.\ D} {\bf 20}, 3-86 (2011).}

\bibitem{isham92}
{C.J.\ Isham, ``Canonical Quantum Gravity and the Problem of Time'', in {\em
  Integrable Systems, quantum Groups, and quantum Field Theories}, eds.\ L.A.\
  Ibort and M.A.\ Rodriguez, Kluwer Academic Publishers, London, 157 (1993) and
  arXiv:gr-qc/9210011.}

\bibitem{anderson17}
{E.\ Anderson, {\em The Problem of Time}, Springer (2017).}

\bibitem{bohm93}
{D.\ Bohm and B.J.\ Hiley, {\em The Undivided Universe}, Routledge, New York
  (1993).}

\bibitem{holland93b}
{P.R.\ Holland, {\em The Quantum Theory of Motion}, Cambridge University Press,
  Cambridge (1993).}

\bibitem{duerr09}
{D.\ D\"urr and S.\ Teufel, {\em Bohmian Mechanics}, Springer-Verlag, Berlin
  (2009).}

\bibitem{goldstein04}
{S.\ Goldstein and S.\ Teufel, ``Quantum spacetime without observers:
  ontological clarity and the conceptual foundations of quantum gravity'', in
  {\em Physics Meets Philosophy at the Planck Scale}, eds.\ C.\ Callender and
  N.\ Huggett, Cambridge University Press, Cambridge, 275-289 (2004) and
  arXiv:quant-ph/9902018.}

\bibitem{pinto-neto05a}
{N.\ Pinto-Neto, ``The Bohm Interpretation of Quantum Cosmology'', {\em Found.\
  Phys.}\ {\bf 35}, 577-603 (2005) and arXiv:gr-qc/0410117.}

\bibitem{pinto-neto19}
{N.\ Pinto-Neto and W.\ Struyve, ``Bohmian quantum gravity and cosmology'', in
  {\em Applied Bohmian Mechanics: From Nanoscale Systems to Cosmology}, 2nd
  edition, eds.\ X.\ Oriols and J.\ Mompart, 607-656 (2019) and
  arXiv:1801.03353 [gr-qc].}

\bibitem{kuchar93}
{K.V.\ Kucha{\u{r}}, ``Canonical Quantum Gravity'', in {\em General Relativity
  and Gravitation 1992}, eds.\ R.J.\ Gleiser, C.N.\ Kozamah and O.M.\ Moreschi,
  Institute of Physics Publishing, Bristol, 119-150 (1993) and
  arXiv:gr-qc/9304012.}

\bibitem{ruffini95}
{G.\ Ruffini, ``Quantization of Simple Parametrized Systems'', PhD.\ thesis, UC
  Davis (1995) and arXiv:gr-qc/0511088.}

\bibitem{rovelli14}
{C.\ Rovelli and F.\ Vidotto, {\em Covariant Loop Quantum Gravity}, Cambridge
  University Press, Cambridge (2014).}

\bibitem{earman02a}
{J.\ Earman, ``Thoroughly Modern McTaggart: Or What McTaggart Would Have Said
  If He Had Learned the General Theory of Relativity'', {\em Philosophers'
  Imprint} {\bf 2}, No.\ 3 (2002).}

\bibitem{maudlin02b}
{T.\ Maudlin, ``Thoroughly Muddled McTaggart: Or How to Abuse Gauge Freedom to
  Generate Metaphysical Monstrosities'', {\em Philosophers' Imprint} {\bf 2},
  No.\ 4 (2002).}

\bibitem{healey02}
{R.\ Healey, ``Can Physics Coherently Deny the Reality of Time?'', {\em R.\
  Inst.\ Philos.\ suppl.}\ {\bf 50}, 293-316 (2002).}

\bibitem{sundermeyer82}
{K.\ Sundermeyer, {\em Constrained Dynamics}, Lecture Notes in Physics 169,
  Springer-Verlag, Berlin (1982).}

\bibitem{henneaux92}
{M.\ Henneaux and C.\ Teitelboim, {\em Quantization of Gauge Systems},
  Princeton University Press, New Jersey (1992).}

\bibitem{struyve09a}
{W.\ Struyve and A.\ Valentini, ``de Broglie-Bohm guidance equations for
  arbitrary Hamiltonians'' {\em J.\ Phys.\ A} {\bf 42}, 035301 (2009) and
  arXiv:0808.0290v3 [quant-ph].}

\bibitem{struyve16}
{A.\ Cesa, J.\ Martin and W.\ Struyve, ``Chaotic Bohmian trajectories for
  stationary states'', {\em J.\ Phys.\ A: Math.\ Theor.}\ {\bf 49}, 395301
  (2016) and arXiv:arXiv:1603.01387 [quant-ph].}

\bibitem{duerr92a}
{D.\ D\"urr, S.\ Goldstein and N.\ Zangh\`\i, ``Quantum Equilibrium and the
  Origin of Absolute Uncertainty'', {\em J.\ Stat.\ Phys.}\ {\bf 67}, 843-907
  (1992) and arXiv:quant-ph/0308039. Reprinted in \cite{duerr12}.}

\bibitem{duerr20b}
{D.\ D\"urr and W.\ Struyve, ``Quantum Einstein equations'', {\em Class.\
  Quantum Grav.}\ {\bf 37}, 135002 (2020) and arXiv:2003.03839 [gr-qc].}

\bibitem{duerr14a}
{D.\ D\"urr, S.\ Goldstein, T.\ Norsen, W.\ Struyve and N.\ Zangh\`\i, ``Can
  Bohmian mechanics be made relativistic?'', {\em Proc.\ R.\ Soc.\ A} {\bf
  470}, 20130699 (2014) and arXiv:1307.1714 [quant-ph].}

\bibitem{vink92}
{J.C.\ Vink, ``Quantum potential interpretation of the wave function of the
  universe'', {\em Nucl.\ Phys.\ B} {\bf 369}, 707-728 (1992).}

\bibitem{squires92b}
{E.J.\ Squires, ``An apparent conflict between the de Broglie-Bohm model and
  orthodoxy in quantum cosmology'', {\em Found.\ Phys.\ Lett.}\ {\bf 5}, 71-75
  (1992).}

\bibitem{shtanov96}
{Y.V.\ Shtanov, ``Pilot wave quantum cosmology'', {\em Phys.\ Rev.\ D} {\bf
  54}, 2564-2570 (1996) and arXiv:gr-qc/9503005.}

\bibitem{callender96}
{C.\ Callender and R.\ Weingard, ``Time, Bohm's Theory, and Quantum
  Cosmology'', {\em Philos.\ Sci.}\ {\bf 63}, 470-474 (1996).}

\bibitem{chua21}
{E.Y.S.\ Chua and C.\ Callender, ``No Time for Time from No-Time'', {\em
  Philos.\ Sci.}\ {\bf 88}, 1172-1184 (2021).}

\bibitem{padmanabhan90}
{T.\ Padmanabhan, ``A definition for time in quantum cosmology'', {\em Pramana
  J.\ Phys}.\ {\bf 35}, L199-L204 (1990).}

\bibitem{greensite90}
{J.\ Greensite, ``Time and probability in quantum cosmology'', {\em Nucl.\
  Phys.\ B} {\bf 342}, 409-429 (1990).}

\bibitem{greensite91}
{J.\ Greensite, ``Ehrenfest's principle in quantum gravity'', {\em Nucl.\
  Phys.\ B} {\bf 351}, 749-766 (1991).}

\bibitem{page83}
{D.N.\ Page and W.K.\ Wootters, ``Evolution without evolution: Dynamics
  described by stationary observables'', {\em Phys.\ Rev.\ D}\ {\bf 27},
  2885-2892 (1982).}

\bibitem{bell82}
{J.S.\ Bell, ``On the Impossible Pilot Wave'', {\em Found.\ Phys.}\ {\bf 12},
  989-999 (1982), reprinted in J.S.\ Bell, {\em Speakable and unspeakable in
  quantum mechanics}, Cambridge University Press, Cambridge (1987).}

\bibitem{thiemann07}
{T.\ Thiemann, {\em Modern Canonical Quantum General Relativity}, Cambridge
  University Press, Cambridge (2007).}

\bibitem{struyve17b}
{W.\ Struyve, ``Loop quantum cosmology and singularities'', {\em Sci.\ Rep.}\
  {\bf 7}, 8161 (2017) and arXiv:1703.10274 [gr-qc].}

\bibitem{duerr12}
{D.\ D\"urr, S.\ Goldstein and N.\ Zangh\`\i, {\em Quantum Physics Without
  Quantum Philosophy}, Springer-Verlag, Berlin (2012).}

\end{thebibliography}
\end{document}